\documentstyle[aps,multicol,psfig,epsfig]{revtex}
\begin{document}
\def\pa{\parallel}
\def\pe{\bot}
\def\xx{{\bf{x}}}
\draft
\tightenlines
 
\title{Temporally disordered Ising models}
\author{
 Juan Jos\'e Alonso$^{1}$ and Miguel A. Mu\~noz$^{2}$}
\address{
$^1$  Departamento de F{\'\i}sica Aplicada 1, Facultad de Ciencias, 
29071 M\'alaga, Spain \\ 
$^2$ Institute {\em Carlos I} for Theoretical and Computational Physics\\
and Departamento de E. y F{\'\i}sica de la Materia,
Universidad de Granada
 18071 Granada, Spain.\\    
}
\date{\today}

\maketitle
\begin{abstract}
 We present a 
 study of the influence of different types of disorder 
on systems in the Ising universality class by employing both
a dynamical field theory approach and extensive Monte Carlo
simulations. 
We reproduce some well 
known results for the case of quenched disorder (random temperature
and random field), and analyze the effect of four different types
of time-dependent disorder scarcely studied so far in the literature.
Some of them are of obvious experimental and theoretical relevance
(as for example, globally fluctuating temperatures or random fields).
 All the predictions coming from our field theoretical analysis 
are fully confirmed by extensive 
simulations in two and three dimensions, and novel
 qualitatively different, non-Ising transitions are reported.
  Possible experimental setups designed to explore the      
  described phenomenologies are also briefly discussed.
\end{abstract}
\begin{multicols}{2}
\narrowtext

 The Ising model has played a central role in the 
development of modern statistical mechanics.
Its universality class  includes a broad variety of 
models and real systems, as magnetic and
 reaction-diffusion systems,  
spin glasses, neural networks, 
coupled maps, and protein folding models \cite{Stanley,Marro}.
 The analysis of the robustness of this universality class upon 
 the introduction of disorder is of outmost importance
 in many different contexts. In fact,
there are plenty of studies
devoted to determining whether, and in which way, quenched
 disorder is a relevant
perturbation at the Ising renormalization group (RG) fixed point 
\cite{Stinch,Geoff,Cardy}. 
   While a lot of attention has been paid to the study of
quenched-in-space disorder \cite{Geoff,Stinch} in its different variants,
quenched-in-time disorder (i.e. homogeneous disorder changing randomly in time)
 has not yet been thoroughly studied
(some exceptions to this assertion are \cite{AchaRF,Hausmann}; see also
\cite{Marro}).
This is surprising as 
stochastically time-changing temperatures,
or magnetic fields, are quite straightforward idealizations of 
realistic situations. 

 Here we study the effect of different types of
disorder on the Ising universality class by 
using field theoretical tools
and extensive Monte Carlo simulations.
 Quenched-in-space disorder will 
be considered for the sake of completeness (in particular,
quenched-in-space temperature (QST), 
and quenched-in-space random field (QSF)), 
 but we will chiefly focus on the
novel and rich phenomenology 
of temporally disordered Ising models, and will analyze extensively the
 following four types of disorder: 
 (1) Quenched-in-time, uniform-in-space, random temperature (QTT).
 (2) Quenched-in-space-time random temperature   (QSTT).      
 (3) Quenched-in-time, uniform-in-space, random (magnetic) field (QTF). 
 (4) Quenched-in-space-time random (magnetic) field (QSTF).
Physically, the difference between the odd and the even cases
 stems from the fact that in the first ones
the fluctuating external parameter couples globally to the system, while
even cases represent situations in which the fluctuating 
parameter is inhomogeneous (changing in space and time).
 In terms on Monte Carlo implementations, 
this difference 
can be enviewed as associated with two different types
 of updating:
if a value of the random variable is selected for each
spin update ({\it sequential} updating),
  one is placed naturally at the second (or fourth) case,
while if the random variable is extracted after updating of
the whole lattice 
(this is the case, for example for {\it parallel} updating)
 the situation is described by the first (or third).
In this way the type of updating can indirectly determine the 
nature of the disorder, and eventually the emerging behavior.

 Our main conclusions are (listed in table 1):
(i) QTT is a relevant perturbation for $d > 2$, but irrelevant in $d=2$.
We find strong numerical evidence of continuously
varying exponents as a function of the width of the
 temperature distribution.
 A qualitatively different phenomenology
is found for strong disorder 
(i.e. for widths above a certain threshold).
(ii) QSTT is always an irrelevant perturbation.
(iii) QTF is relevant for $d > 1$;
 and a phenomenology completely different from the standard Ising one
emerges.
(iv) QSTF is marginal for $d = 4$, and irrelevant otherwise.

 Let us start discussing the disordered-temperature cases. 
 The most basic result for QST disorder is
the Harris criterion \cite{Harris} that
can be sketched as follows: 
If $T$ is a random uncorrelated \cite{uncorrelated}
 quenched-in-space variable, 
the amplitude of its fluctuations per unit volume, $f$, in 
a region of linear size $\xi$ is proportional to 
 $\xi^{-d/2}$. At a distance $\Delta$ of the
critical point and taking $\xi$
 to be the correlation length, one has  $\xi \sim \Delta^{-\nu}$,
and  $ f \propto \Delta^{{\nu d} \over 2}$.            
If ${\nu d}  \leq 2$ temperature fluctuations control 
the critical scaling,
while they are irrelevant otherwise  \cite{alfa}.
To study the QTT disorder, 
one may replace $\xi$  by a characteristic
time $t_0 \sim \Delta^{\nu z}$ 
and follow analogous steps to obtain $\nu z \leq 2$
 as the proper relevance criterion. 

In order to make more rigorous and general the previous arguments 
let us now re-derive them in a field theoretical context \cite{Harris}.
First we cast the Ising model
into a Ginzburg-Landau-Wilson (GLW)
 dynamical functional \cite{BJW}, 
whose associated action is
\begin{eqnarray}
{\cal{S}} &  =  &\int d \xx dt \large\{ {\lambda \over 2} \psi^2(\xx,t)
- \psi(\xx,t) [\dot{\phi}(\xx,t)- \tau \phi -  \nonumber \\
 &&{ \lambda \over 2} \nabla^2
\phi(\xx,t) + {g \over 6} \phi^3 (\xx,t) +  h(\xx,t) ] \large\}
\label{MSR}
\end{eqnarray} 
where $\phi$ is the magnetization field, $\psi$ is the response function, 
$\lambda$ 
 and $g$ are parameters, $\tau  \propto T-T_c$ includes the relevant
dependence on the temperature \cite{Amit},
 and $h$ is a magnetic field.

In order to include the effect of a disordered temperature we
substitute $\tau$ by a function $\tau(\xx,t)$.
Then we can resort to one among two alternatives:
 one
is to employ the ``replica'' trick \cite{Geoff}, while here we follow
the equivalent dynamical approach 
introduced some time ago by De Dominicis \cite{noreplicas}.
In fact, the dynamical approach
is the most natural one for the
study of time-dependent disorder. The next step is to average 
the dynamical generating functional  associated with 
Eq. (\ref{MSR}) over the distribution of disorder. In particular,
writing $\tau(\xx,t)=\tau_0 + \delta \tau(\xx,t)$ 
(where $\tau_0$ is the averaged 
temperature and $\delta \tau(\xx,t)$ is a zero-mean,
$\sigma$ variance, Gaussian 
distributed random variable  \cite{uncorrelated})
 and performing the Gaussian
integration, one obtains an effective action identical to         
 Eq.(\ref{MSR}) except for the replacement $\tau \rightarrow \tau_0$
and for the presence of the following  additional disorder-dependent 
 (nonlocal) terms:
\begin{eqnarray}
\sigma \int d \xx \left[ \int dt \phi(\xx,t) 
\psi(\xx,t) \right]^2  & for ~~~~ QST
\label{QST} \\
 \sigma \int dt \left[\int d \xx \phi(\xx,t)
 \psi(\xx,t) \right]^2   & for ~~~~ QTT
\label{QTT}  \\   
 \sigma  \int d \xx \int dt  \left[ \phi(\xx,t)
 \psi(\xx,t) \right]^2  & for ~~~~ QSTT.
\label{QSTT}      
\end{eqnarray}

Let us analyze separately these three cases. For  
QST the naive dimension of the 
coefficient of the extra (non-Markovian)
term is such that Eq. (\ref{QST}) is dimensionless. As 
$ \int dt \int d{\xx} \tau_0 \phi \psi$ has also to be dimensionless,
$[\int dt \phi \psi]= \Lambda^{d-1/\nu}$. Substituting this into 
the previous expression one obtains $[\sigma] = \Lambda^{2/\nu-d}$,
 and therefore
$\sigma$ is relevant (in the infrared limit \cite{Amit})
if $d \nu \leq 2$ (reproducing the Harris criterion).
 Proceeding analogously 
for QTT one obtains  $\nu z \leq 2$ and 
for QSTT  $ \nu (d +z )  \leq 2$ as relevance criteria.
 Substituting the known Ising exponent values \cite{Amit},
 one finds that:
(see table) QST is a 
marginal perturbation at $d=2$ \cite{Queiroz} 
and relevant at $d=3$,  as already known \cite{Geoff,unico}.        
QTT is relevant for $d \geq 3$, while it is irrelevant 
in $d=2$. QSTT is irrelevant in any dimension. This prediction
implies that homogeneous globally changing temperatures (QTT) are far more
effective in altering Ising behavior than locally changing 
temperatures (QSTT). 

For the case of random fields, we proceed along analogous lines.
By averaging the generating functional over a magnetic-field Gaussian 
distribution (centered at zero and of width $\delta$)
 \cite{uncorrelated}, we obtain the 
terms to be added to Eq. (\ref{MSR}):
(i)   $ \delta \int d \xx
 \left[ \int dt \psi(\xx,t) \right]^2 $ for QSF
(ii) $ \delta \int dt 
\left[\int d \xx \psi(\xx,t) \right]^2$ for QTF, and          
(iii)  $ \delta \int d \xx \int dt  \psi^2(\xx,t)$ and 
for QSTF.

By noticing that the dimension of
$ \int dt  \psi(\xx,t)$ 
at the pure Ising RG fixed point       
in terms of $\Lambda$ is
$ {d \over 2} - {\gamma \over 2 \nu} $
(where $\gamma$ is the susceptibility exponent),
 it is easy to derive the following relevance criteria:
 $\gamma / \nu \geq 0$ for QSF,
 $d \nu + \gamma \geq \nu z$ for QTF, and
 $\gamma \geq \nu z$ for QSTF.
Substituting the  Ising exponents \cite{Amit} we are lead to
the following predictions (see table 1):
 (i) QSF is relevant for $d > 1$. 
 For this case, the Imry-Ma argument establishes
that $d=2$ is the lower critical dimension of the QSF universality class
\cite{unico}).
(ii) QTF is relevant in all dimensions. 
 Some previous results about
 this model can be found 
\cite{AchaRF,Hausmann}; in particular, 
it is expected to
exhibit a transition even in  $d=1$, and  
a non Ising-like 
phase transition has been reported in $d=2$ \cite{Hausmann}. 
(iii) QSTF is marginal only at $d = 4$, and irrelevant otherwise
(in fact, it just changes the noise amplitude, therefore it does not
introduce any new operator). 
Analogously to what happens in the  case of disordered       
temperatures, homogeneous global magnetic fields (QTF) are 
more effective in affecting the universal properties of the pure Ising
class than magnetic fields varying both in time and in space (QSTF).
 
  Let us finally stress that all the relevance criteria are 
 determined using anomalous exponents, therefore
 at the Ising and not at the Gaussian fixed point. 
By performing naive power counting 
one can easily see, that $QTT$ and $QTF$ are also relevant 
perturbations at the Gaussian, mean-field, fixed point.
Hence they are relevant even above the Ising
 upper critical dimension, $d_c=4$.

Having established all these relevance criteria,
the next natural step would be to 
determine the new upper critical dimensions, $d_c$,
 in the cases where disorder
is relevant, and perform an  $\epsilon-$expansion in each case around 
the corresponding $d_c$.  For example, for QSF 
 the extra nonlocal operator
 scales as $\Lambda^2$
at $d=4$ (it is more relevant than the Ising nonlinearities), and
accordingly the
upper critical dimension is shifted from $4$ to $6$. 
To every order in an $\epsilon-$expansion 
the exponents of the QSF model in $d$ dimensions are identical 
to those of the pure problem in $d-2$ dimensions \cite{DR,Geoff}. 

Our attempts to perform similar detailed  RG calculations in the other  
cases have turned out to be extremely hard and, so far, inconclusive
 \cite{Janssen}.
For instance, for QTT and QTF the degree of divergence of the nonlocal
 operator is
insensitive to dimensionality above the
Ising critical dimension $d_c=4$.
 Consequently there is no upper critical dimension for these
 types of disorder which are
 strongly relevant even at high dimensions.
Hence, even finding 
a sound mean-field approach,  
 around which one could perform some sort
of perturbative approach, turns out to be a nontrivial task. 
The only remaining possibility we can think of would be
to perform a 
double expansion analogous to that in \cite{BAW}.        
In any case, at the present stage of our research, we can just 
predict whether disorder is relevant or not, but cannot make 
sensible predictions concerning eventual new RG
fixed points and the nature of new transitions.
   
 In order to test the above predictions, and explore the eventual existence
of new universality classes and/or novel phenomenologies,
we have performed extensive Monte Carlo simulations 
combined with finite size scaling analysis
in $d=2$ and $d=3$ with 
different types of disorder.
 We use: system sizes up to $L=512$ ($L=90$) in
 $d=2$ ($d=3$),
 periodic boundary conditions,
 Metropolis \cite{Marro} transition rates, and run
 up to $3 \times 10^7$ Monte Carlo steps. 
 Here we report the main results,
further details and discussions will be presented elsewhere \cite{Future}:

(i) For disordered temperature cases we have performed 
simulations using both QTT and QSTT. 
The temperature is distributed homogeneously and symmetrically
around a central (tunable)
value $T_0$ and width $\delta$ \cite{uncorrelated}.
 For both types of disorder
in $d=2$ we obtain exponents compatible with the respective systems 
being in the Ising class; see inset in Figure 1, where we
study the scaling of the susceptibility peak 
(maximum value of the magnetization fluctuations) 
 with system size $L$.
 The model studied in \cite{Puma} can also be classified 
as QSTT in $d=2$, and in fact, it was reported to exhibit Ising scaling.
  In $d=3$ for QSTT from the curve $\chi$ vs $L$ we observe very good 
scaling using the three-dimensional $\gamma/\nu$ Ising value.
On the other hand, for QTT we observe
 exponents varying continously with $\delta$, changing from the Ising value
$\gamma/\nu =1.97(1)$ for $\delta=0$, to $\gamma/ \nu=3$ for values
of $\delta \approx 1.5$. There is no evidence of
crossovers and the variation in exponent values is rather 
broad. Therefore we believe that these are the real 
asymptotic values and that accordingly there should be a line of RG
 fixed points. Caution should be used before concluding that
 the exponents actually
change continuously. Corrections to scaling
could be responsible for such a variation                    
 \cite{caution}. 
 For  values of $\delta$ larger than a certain
threshold around $1.5$ the exponent $\gamma/ \nu$ saturates to $3$
implying that the magnetization histograms do not converge
to delta Dirac functions in the thermodynamic limit: they remain 
broad, corresponding to external-fluctuations-controlled transitions.

(ii) For the case of random fields, QTF is        
expected to be non-Ising in $d=2$, while QSTF disorder is predicted to be
an irrelevant perturbation.
 In fact, for QSTF our simulations show a perfect agreement 
with Ising exponents for all the measured magnitudes; non-universal
properties depend on the disorder width, $\delta$,
 but critical exponents remain 
unchanged. The models studied in \cite{ALM,Marro} can also be 
ascribed to this class of disorder, and were found to be Ising like 
in $d=2$ in accordance with our prediction.
 On the other hand for QTF we observe a completely 
different phenomenology: There is a continuous phase transition
whose features are unambiguously different from the 
corresponding Ising ones.
In particular, it is a sort of
dynamical transition controlled by the fluctuations of the
randomly changing magnetic field. 
For example, in figure 2, we show the dependence of the susceptibility
per unit surface as a function of system size, for different
values of $\delta$. After a finite-size-effect transient, the
curves converge to a fixed value, indicating the existence of
broad magnetization histograms in the thermodynamic limit,
and implying $\gamma/\nu=2$.
Still, magnitudes
as the specific heat or the susceptibility show singular
peaks at the critical point, and there is a dynamical spontaneous  
symmetry breaking below the critical temperature \cite{Future,bimodal}.
We have reproduced qualitatively this QTF phenomenology by means
of a dynamical mean-field approach
to be reported elsewhere \cite{Future}, and observed that
the nature of this transition does not seem to be very sensible
to dimensionality changes.

  Summing up: we have presented a rather simple general field theoretical 
scheme allowing  us to predict the relevance of different types of 
disorder in the Ising universality class. Some of the conclusions 
where already known, and provide a validity test for our approach. 
 Using this scheme we have studied the effect of
 different types of time-dependent
disorder in the Ising class. These disordered models describe
 rather natural situations
in which systems are subjected to the influence
of some globally (or locally) fluctuating external agent (randomly
varying temperatures or magnetic fields).  
All our predictions are fully confirmed by means of extensive Monte Carlo 
simulations; and novel quite interesting phenomenologies appear
for the case of homogeneously fluctuating temperatures and
magnetic fields.     

In order to experimentally verify our predictions 
one could take, for example, vacuum deposited ultrathin 
Fe films on W(110) (i.e. a nearly ideal
two-dimensional magnetic Ising system \cite{experiments}), and
in the same way that this material has been set 
to study hysteresis in the presence
of an oscillating magnetic field \cite{experiments},  
it can be analyzed under the effect of a randomly varying
magnetic field (generated, for example, by means of an electric
circuit in a chaotic regime). 
A detailed comparison between the characteristic relaxation
time-scale, and the typical magnetic field variation 
scale is of outmost importance in order to determine
 the applicability 
of our results to this type of experiments.

{\it ACKNOWLEDGMENTS-}
 It is a pleasure to acknowledge J. Marro, P. L. Garrido,
and J.J. Ruiz-Lorenzo
for useful discussions and criticisms.
This work has been partially supported by the
European Network Contract ERBFMRXCT980183
and by the Ministerio de Educaci\'on:
 projects DGESEIC PB97-0842 and PB97-1080.

\begin{figure}
\centerline{
\epsfxsize=3in
\epsfbox{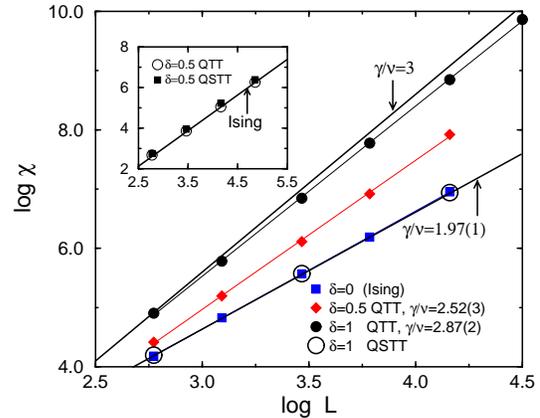}
}
\caption{Susceptibility at the critical point
 as a function of system size for
disordered-temperature models in $d=3$ ($d=2$ in the inset).
 QSTT exhibits Ising scaling in 
$d=3$. For QTT the slopes 
(fixing $\gamma/\nu$) change
continuously with disorder amplitude  
in $d=3$, but not in $d=2$ (see inset).
}
\label{Figure1}
\end{figure}
\begin{figure}
\centerline{
\epsfxsize=3in
\epsfbox{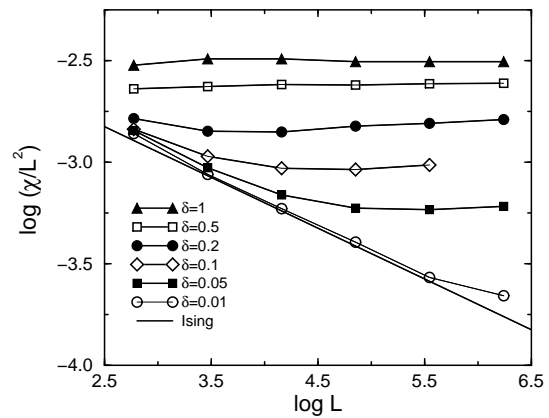}
}
\caption{Susceptibility per unit surface as a function of L
(in log-log scale) for QTF disorder.
For large sizes the curves converge to a constant
indicating the existence of broad
magnetization histograms in the thermodynamic limit.
}
\label{Figure2}
\end{figure}

\begin{table}
\begin{center}
\begin{tabular}{|c|c|c|c|}
$ DISORDER $  & $ Space $ &  $ Time $   & $ Space-Time $  \\
\hline
\hline
           &    $QST$         &    $QTT$        &   $QSTT$   \\
$ Random$  & $ d \nu \leq 2$  & $ \nu z \leq 2 $ & ${
  \nu (d+ z) \leq 2}$ \\
$Temperature$ & $ d \geq 2 $  & $   d  \geq 3  $ &  ${ --- }$   \\
\hline
              &    $QSF$         &    $QTF$        &   $QSTF$   \\     
$ Random $ & $  \gamma/\nu \geq 0   $  & ${ d \nu +\gamma \geq \nu z }$ &
 $ { \gamma > \nu z } $ \\
$Field $   &  $d  > 1  $ &  $ d \geq 1$ &  $ d = 4 ~~ marginal$ \\
\end{tabular}
\end{center}
\label{tabla}
\caption{\small
 Criteria for the relevance of different types of disorder
in the Ising universality class.
The first row in each case specifies the general
criteria (equalities correspond to marginal cases),
while in the second we specify the dimensions at which such criteria
are fulfilled.
}
\end{table}                
\end{multicols}

\begin{thebibliography}{99}
\bibitem{uncorrelated} As long as the disorder at different
sites is uncorrelated, details of the one-site distribution
are irrelevant.
\bibitem{alfa} Using the scaling law $\nu d = 2- \alpha$
\cite{Amit}
the relevance
criterion can be written in its more familiar form, $\alpha > 0$.
\bibitem{bimodal}  A similar situation was studied in \cite{Hausmann}:
 the disorder distribution 
was bimodal there, and this made
the transition to be a first-order one.
\end{thebibliography}
\end{document}